\documentclass[aps,prl,reprint,groupedaddress]{revtex4-1}
\usepackage{graphicx}
\usepackage{color}
\newcommand{\beq}{\begin{equation}}
\newcommand{\eeq}{\end{equation}}
\newcommand{\bea}{\begin{eqnarray}}
\newcommand{\eea}{\end{eqnarray}}
\begin{document}

%\preprint{}

\title{Freeze-out parameters from electric charge and baryon number fluctuations: \\is there consistency?}


\author{S. Borsanyi$^{1}$, Z. Fodor$^{1,2,3}$, S. D. Katz$^{2,4}$, S. Krieg$^{1,3}$, C. Ratti$^{5}$, K. K. Szabo$^{1,3}$}
%\thanks{}
\affiliation{$^1$ \small{\it Department of Physics, Wuppertal University, Gaussstr. 20, D-42119 Wuppertal, Germany}\\
$^2$ \small{\it Inst. for Theoretical Physics, E\"otv\"os University,}\\
\small{\it P\'azm\'any P. s\'et\'any 1/A, H-1117 Budapest, Hungary}\\
$^3$ \small{\it J\"ulich Supercomputing Centre, Forschungszentrum J\"ulich, D-52425
J\"ulich, Germany}\\
$^4$ \small{\it MTA-ELTE "Lend\"ulet" Lattice Gauge Theory Research Group,}\\
\small{\it P\'azm\'any P. s\'et\'any 1/A, H-1117 Budapest, Hungary}\\
$^5$ \small{\it Dip. di Fisica, Universit\`a di Torino and INFN, Sezione di Torino}\\
\small{\it via Giuria 1, I-10125 Torino, Italy}\\}

\date{\today}

\begin{abstract}
Recent results for moments of multiplicity distributions of net-protons and net-electric charge from the STAR collaboration are compared to lattice QCD results for higher order fluctuations of baryon number and electric charge by the Wuppertal-Budapest collaboration, with the purpose of extracting the freeze-out temperature and chemical potential. All lattice simulations are performed for a system of 2+1 dynamical quark flavors, at the physical mass for light and strange quarks; all results are continuum extrapolated. We show that it is possible to extract an upper value for the freeze-out temperature, as well as precise baryo-chemical potential values corresponding to the four highest collision energies of the experimental beam energy scan. Consistency between the freeze-out parameters obtained from baryon number and electric charge fluctuations is found. The freeze-out chemical potentials are now in agreement with the statistical hadronization model.

\end{abstract}
\pacs{}
%\keywords{}

\maketitle

The fundamental theory of strong interactions predicts a transition from a hadronic system
to a partonic one at sufficiently high temperatures or densities;
this transition is an analytic crossover, as shown
by lattice QCD simulations \cite{Aoki:2006we}. The conditions of temperature or density needed
to create the deconfined phase of QCD can be reached in the laboratory: 
relativistic heavy ion collisions are contributing tremendously to our understanding of the 
QCD phase diagram and the properties of the quark-gluon plasma (QGP).
The precision reached in the most recent lattice QCD simulations, as well as the
increasing availability of data from the heavy ion experimental programs,
allow us to compare the theoretical and experimental results in a very efficient way.

Fluctuations of conserved charges (electric charge, baryon number and strangeness) are
certainly a major example of this fruitful synergy. These observables, especially the
higher order moments, were originally proposed as a signature for the
QCD critical point \cite{Stephanov:1999zu,Gavai:2008zr,Cheng:2007jq}, 
namely the point of the phase diagram which marks the
change from crossover to first-order phase transition. 
As a consequence, experimental results for moments of net-electric 
charge and net-proton multiplicity distributions have
recently been published, in the collision-energy range $\sqrt{s}=7.7-200$ GeV covered by
the RHIC beam energy scan
\cite{McDonald:2012ts,Sahoo:2012bs,Adamczyk:2013dal,Adamczyk:2014fia}.
The recent idea of extracting the freeze-out parameters of a heavy-ion collision
through a comparison between lattice QCD results and experimental data has
renewed the interest towards these observables even at $\mu=0$
\cite{Karsch:2012wm,Bazavov:2012vg,Borsanyi:2013hza}.

The chemical freeze-out, defined as the last inelastic scattering of hadrons before detection, has
already been studied in terms of the statistical hadronization model by fitting a chemical potential
and a temperature parameter to the pion, kaon, proton and other accessible yields from experiment
\cite{Andronic:2008gu,Becattini:2005xt,Cleymans:2005xv,Manninen:2008mg}.
By decreasing the collision energy, the freeze-out chemical potential increases; repeating
the analysis for a series of beam energies provides a freeze-out curve in the
$(T,\mu)$ plane.

The comparison between the experimental and lattice QCD results for the 
electric charge and baryon number fluctuations
allows a first-principle determination of the freeze-out temperature and chemical potential, under the 
assumption that the experimentally measured fluctuations can be described in terms
of the equilibrium system simulated on the lattice. Possible experimental sources of non-thermal fluctuations
are corrected for in the STAR data analysis: the centrality-bin-width correction method minimizes effects due to volume variation because of finite centrality bin width; the moments are corrected for the finite reconstruction efficiency based on binomial probability distribution \cite{Bzdak:2012ab}; the spallation protons are removed by appropriate cuts in $p_T$ \cite{Adamczyk:2014fia}. In general, effects due to kinematic cuts are found to be small \cite{Alba:2014eba}.
Besides, final-state interactions in the hadronic phase and non-equilibrium effects might become
relevant and affect fluctuations \cite{Steinheimer:2012rd,Becattini:2012xb}, therefore a fundamental check in favor of the equilibrium scenario is the consistency between the freeze-out parameters yielded by different quantum numbers e.g. electric charge and baryon number. In particular, while the freeze-out temperature might be flavor-dependent \cite{Bellwied:2013cta}, the chemical potentials as a function of the collision energy should be the same for all species. The present level of precision reached by lattice QCD results,
obtained at physical quark masses and continuum-extrapolated, allows to perform this check for the first time.

One more caveat is in order, since experimentally only the net-proton multiplicity distribution is measured, as opposed to the lattice net-baryon number fluctuations. 
%Due to the fact that isospin-fluctuations remain finite at the critical
%point, the critical fluctuations in the net-baryon number
%are directly imprinted in the net-proton fluctuations \cite{Hatta:2003wn}.
%Sources of finite and non-equilibrium fluctuations can,
%however, significantly hide the critical fluctuations in the
%net-proton number as compared to net-baryon number
%fluctuations \cite{Kitazawa:2011wh,Kitazawa:2012at}. 
Recently it was shown that,
once the effects of resonance feed-down and isospin randomization are 
taken into account \cite{Kitazawa:2011wh,Kitazawa:2012at}, the net-proton and net-baryon number fluctuations are numerically 
very similar \cite{Nahrgang:2014fza}.

In this paper we show for the first time that it is possible to find a 
consistency between the freeze-out parameters yielded by electric charge and
baryon number fluctuations. This is achieved by systematically comparing our
continuum-extrapolated results for higher order
fluctuations of these conserved charges \cite{Borsanyi:2013hza} 
to the corresponding experimental data by the STAR collaboration at RHIC
\cite{Adamczyk:2013dal,Adamczyk:2014fia}.  We are using the newly
published, efficiency-corrected experimental results for the net-charge
fluctuations and combine them with our lattice results presented in
Ref.~\cite{Borsanyi:2013hza}. We also extract independent freeze-out
conditions from the net-proton fluctuations and systematically compare the
outcomes of the two.  Details of the lattice simulations can be found 
in \cite{Borsanyi:2013hza}.

The fluctuations of baryon number, electric charge and strangeness are defined as
\bea
\chi_{lmn}^{BSQ}=\frac{\partial^{\,l+m+n}(p/T^4)}{\partial(\mu_{B}/T)^{l}\partial(\mu_{S}/T)^{m}\partial(\mu_{Q}/T)^{n}};
\eea
they are related to the moments of the multiplicity distributions of the corresponding conserved charges. It is convenient to introduce the following, volume-independent ratios
\bea
\chi_3/\chi_{2}=~S\sigma
\quad&;&\quad
\chi_4/\chi_{2}=\kappa\sigma^2
\nonumber\\
\chi_1/\chi_2=M/\sigma^2
\quad&;&\quad
\chi_3/\chi_1=S\sigma^3/M\,.
\label{moments}
\eea

The chemical potentials
$\mu_B,~\mu_Q$ and $\mu_S$ are related in order to match
the experimental situation: the finite baryon density in the system is due to light quarks only, since it is generated by the nucleon stopping in the collision
region. The strangeness density $\langle n_S\rangle$ is then equal to zero for all collision energies, as a consequence of strangeness conservation. Besides, the electric charge and baryon-number densities are related, in order to match the isospin asymmetry of the colliding nuclei: 
\hbox{$\langle n_Q\rangle=Z/A\langle n_B\rangle$}.
$Z/A=0.4$ represents a good approximation for Pb-Pb and Au-Au
collisions.
%
%\hbox{$\langle n_Q\rangle=0.4\langle n_B\rangle$}.
%For Au-Au and Pb-Pb
%collisions, a good approximation is to assume 
%\hbox{$\langle n_Q\rangle=0.4\langle n_B\rangle$}.

\begin{figure}[b]
%\begin{minipage}{0.48\textwidth}
 \scalebox{.7}{
 \includegraphics{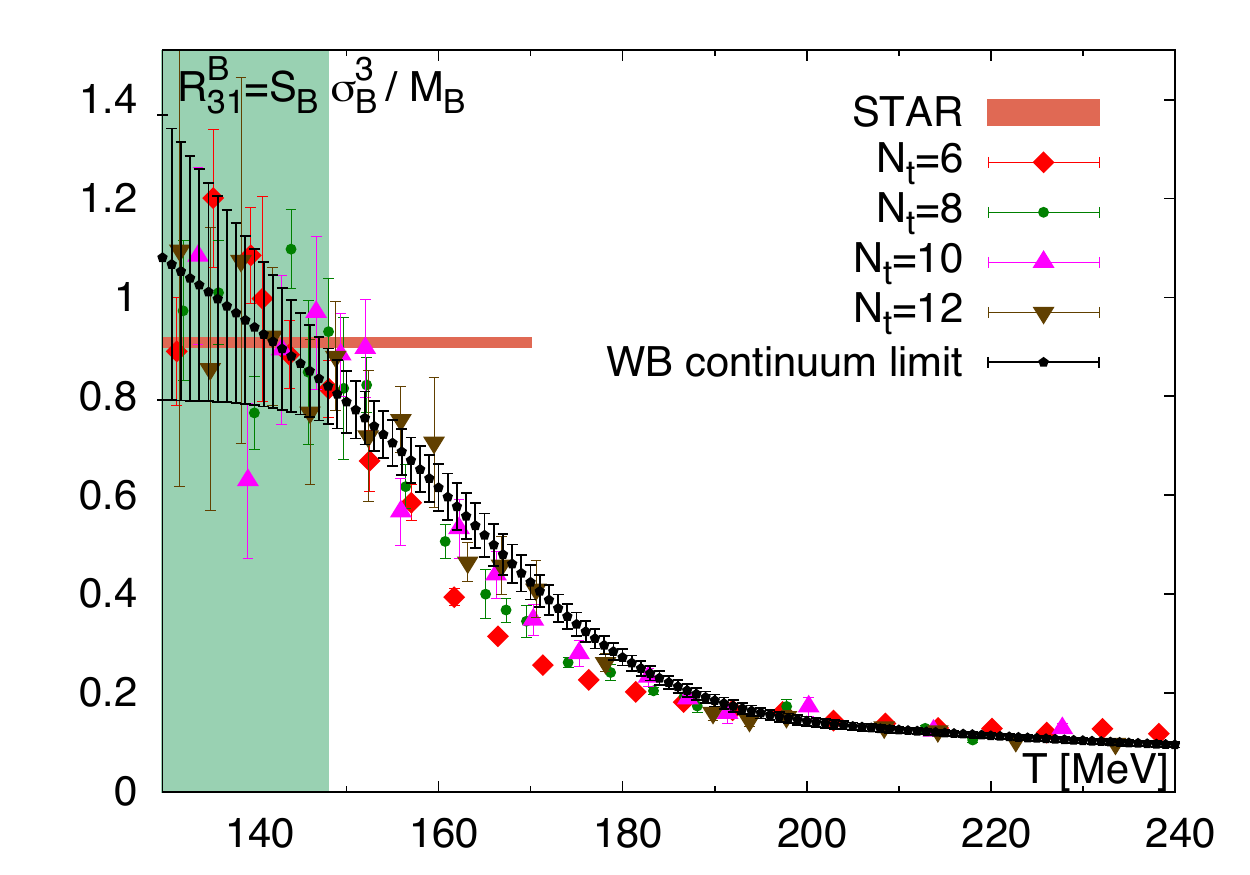}}
% \end{minipage}
%\begin{minipage}{0.48\textwidth}
% \scalebox{.5}{
% \includegraphics{muq}}
% \end{minipage}
\caption{$R_{31}^B$: the colored symbols show finite-$N_t$ lattice QCD results.
 The continuum extrapolation is represented by black points (from Ref. \cite{Borsanyi:2013hza}). 
 The dark-orange band shows the recent
 experimental measurement by the STAR collaboration \cite{Adamczyk:2013dal}: it was obtained by averaging the 0-5\% and 5-10\% data at the four highest energies ($\sqrt{s}=27,~39,~62.4,~200$ GeV). The green-shaded area shows the valid temperature range. \label{fig1}}
\end{figure}
As a consequence, $\mu_Q$ and $\mu_S$ depend on $\mu_B$ 
so that these conditions are satisfied. This is achieved by
Taylor-expanding the densities in these three chemical potentials up
to $\mu_B^3$ \cite{Bazavov:2012vg}:
\bea
\mu_Q(T,\mu_B)&=&q_1(T)\mu_B+q_3(T)\mu_B^3+...
\nonumber\\
\mu_S(T,\mu_B)&=&s_1(T)\mu_B+s_3(T)\mu_B^3+...
\label{eq:q1q2}
\eea
Our continuum extrapolated results for the functions 
$q_1(T),~q_3(T),~s_1(T),~s_3(T)$ were shown
in \cite{Borsanyi:2013hza}.
The quantities that we consider to extract the freeze-out $T$ and $\mu_B$, are the ratios $R_{31}^B=\chi_3^B/\chi_{1}^B$ and
$R_{12}^B=\chi_1^B/\chi_{2}^B$ respectively, at values of $(\mu_B,\mu_Q,\mu_S)$, which satisfy the pyhsical conditions discussed in the previous paragraph. 
%The first terms of their Taylor expansion around $\mu_B=0$
%read:
%\bea
%&&R_{31}^B(T,\mu_B)=\frac{\chi_3^B(T,\mu_B)}{\chi_1^B(T,\mu_B)}=
%\nonumber\\
%&&\frac{\chi_{4}^{B}(T,0)+\chi_{31}^{BQ}(T,0)q_1(T)+\chi_{31}^{BS}(T,0)s_1(T)}{\chi_{2}^{B}(T,0)+\chi_{11}^{BQ}(T,0)q_1(T)+\chi_{11}^{BS}(T,0)s_1(T)}+\mathcal{O}(\mu_B^2)
%\nonumber\\
%&&
%\nonumber\\
%&&R_{12}^B(T,\mu_B)=\frac{\chi_1^B(T,\mu_B)}{\chi_2^B(T,\mu_B)}=
%\nonumber\\
%&&\frac{\chi_{2}^{B}(T,0)+\chi_{11}^{BQ}(T,0)q_1(T)+\chi_{11}^{BS}(T,0)s_1(T)}{\chi_{2}^{B}(T,0)}\frac{\mu_B}{T}
%+\mathcal{O}(\mu_B^3).
%\nonumber\\
%&&
%\label{LO}
%\eea
As shown in Ref.  \cite{Borsanyi:2013hza}, the leading order in $\chi_3^B/\chi_1^B$ is independent of $\mu_B$, while the LO in $\chi_2^B/\chi_1^B$ is linear in $\mu_B$. This allows us to use $R_{31}^B$ to extract the freeze-out temperature; the ratio $R_{12}^B$ is then used to extract $\mu_B$ (notice that our results for $R_{12}^B$ are obtained up to NLO in $\mu_B$).

We then independently extract $\mu_B$ from $\chi_1^Q/\chi_{2}^Q$ (which is also linear in $\mu_B$ to LO), in order to check whether different conserved charges yield consistent freeze-out parameters. In Ref. \cite{Borsanyi:2013hza}, we compared the lattice results for $\chi_3^Q/\chi_{1}^Q$ to the preliminary, efficiency-uncorrected data from the STAR collaboration, to extract an upper limit for the freeze-out temperature. We then obtained the corresponding chemical potentials by performing the same kind of comparison for $\chi_1^Q/\chi_{2}^Q$. The new, efficiency-corrected results for the moments of the net-charge multiplicity distribution
from STAR show significant differences, compared to the uncorrected ones. This yields different values for $\mu_B$, compared to the ones obtained in \cite{Borsanyi:2013hza}. As for $\chi_3^Q/\chi_{1}^Q$, the
experimental uncertainty on the corrected data is such that presently it is not possible to extract a meaningful freeze-out temperature from this observable. 

In Fig. \ref{fig1} we show the comparison between the lattice results for $\frac{\chi_3^B(T,\mu_B)}{\chi_1^B(T,\mu_B)}$ and the experimental measurement of $S_p\sigma_p^3/M_p$ by the STAR collaboration  \cite{Adamczyk:2013dal}. The latter has been obtained for a 0-10\% centrality, at the four highest energies ($\sqrt{s}=27,~39,~62.4,~200$ GeV). Since the curvature of
the phase diagram is small around $\mu_B=0$ \cite{Endrodi:2011gv}, this average
allows to determine the freeze-out temperature. The green-shaded area shows the valid temperature range: due to the uncertainty on the lattice results in the low-temperature regime, it is only possible to extract an upper value for the freeze-out temperature: the freeze-out takes place at 
a temperature $T_f\lesssim148$ MeV, which is somewhat lower than expected from previous analyses
\cite{Mukherjee:2013lsa}
(allowing for a two-sigma deviation for the lattice simulations and the experimental measurements, the highest possible $T_f$ is 151 MeV). In Refs.~\cite{Aoki:2009sc,Borsanyi:2010bp} we have published the lattice
determination of the transition
temperature from various chiral observables in the range 147-157 MeV. For the
minimum of the speed of sound
we found 145(5) MeV in \cite{Borsanyi:2010cj}. The discussed freeze-out temperature
is thus in the cross-over region
around or slightly below the central value.

\begin{figure}[h]
%\begin{minipage}{0.48\textwidth}
 \scalebox{.7}{
 \includegraphics{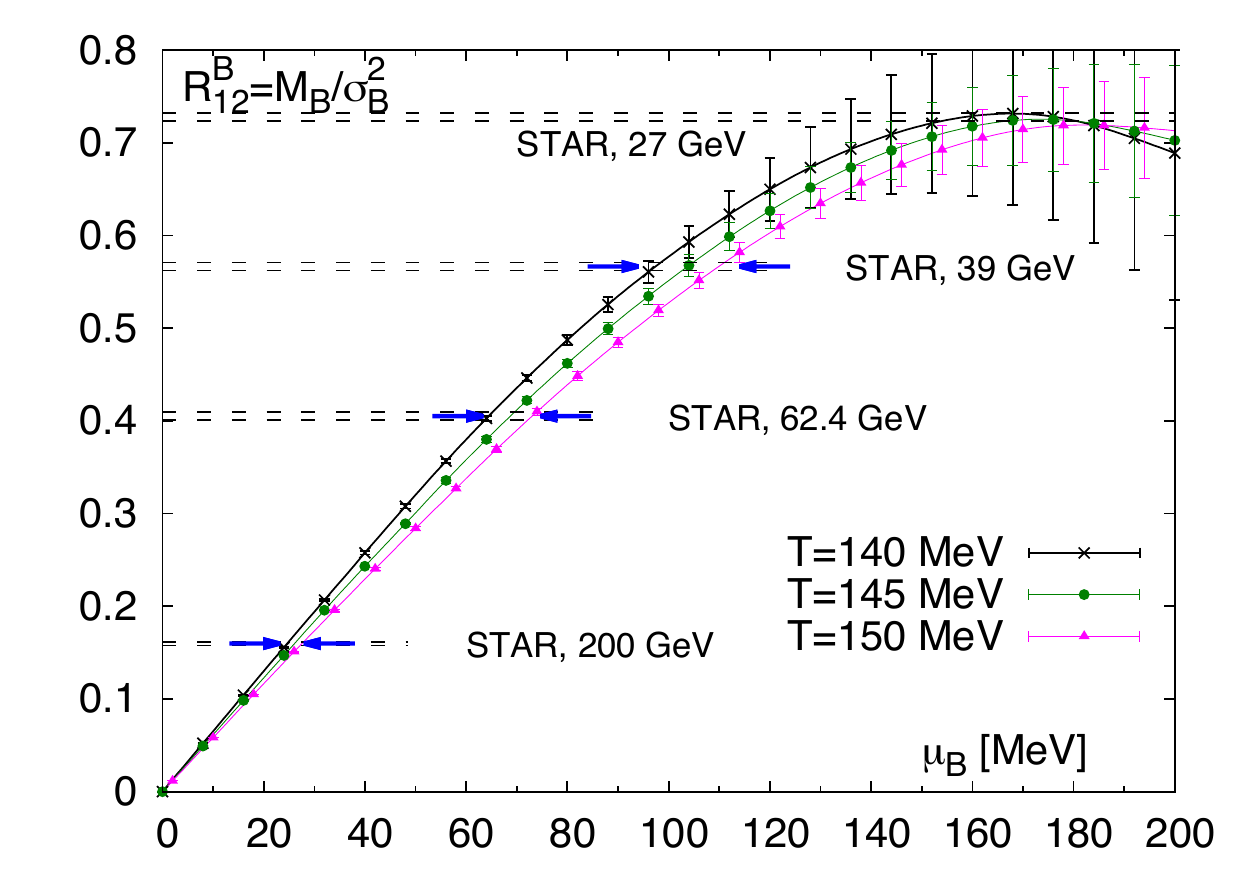}}
% \end{minipage}
%\begin{minipage}{0.48\textwidth}
 \scalebox{.7}{
 \includegraphics{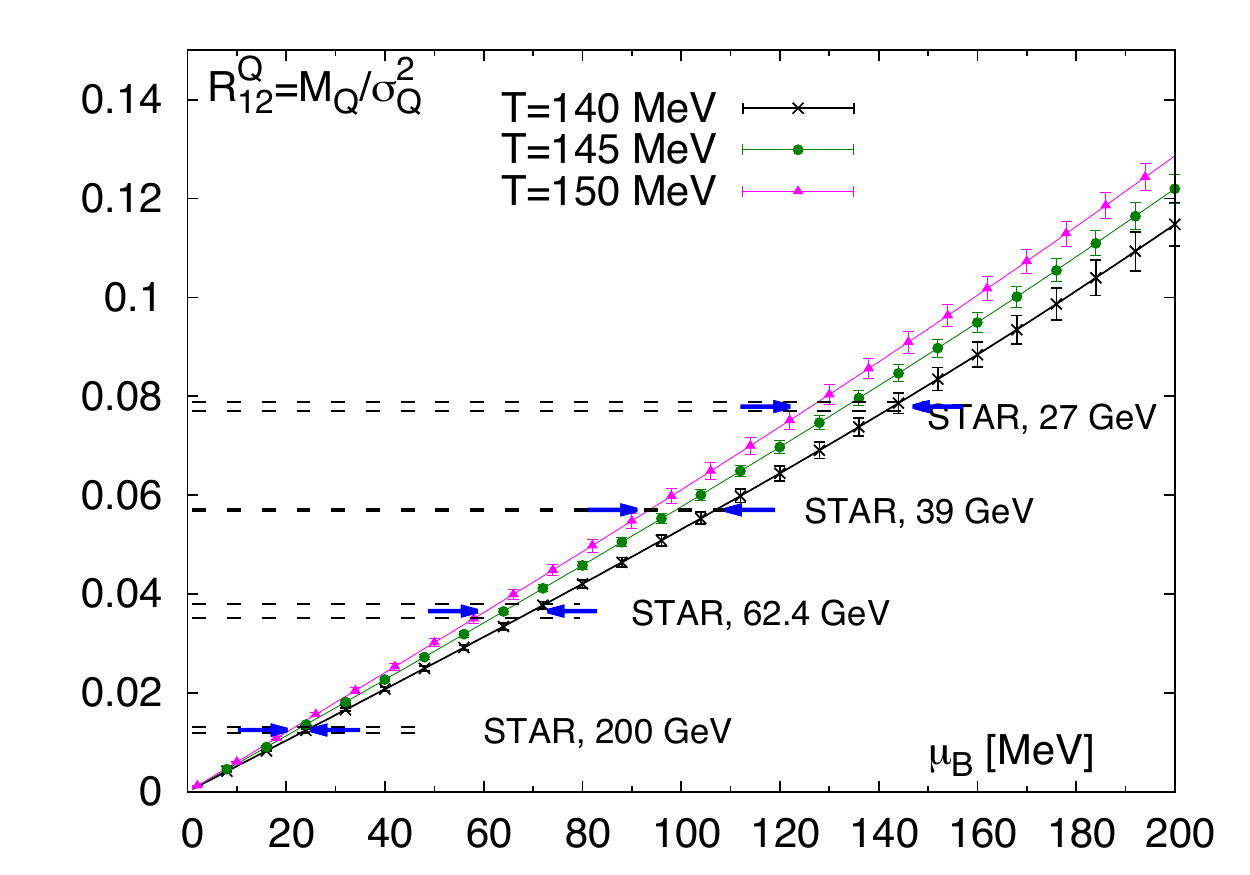}}
% \end{minipage}
\caption{Upper panel: $R_{12}^B$ as a function of $\mu_B$. The three points correspond to the STAR data for $M_p/\sigma_p^2$ at collision energies  $\sqrt{s}=39,~62.4,~200$ GeV and centrality 0-10\%, from Ref. \cite{Adamczyk:2013dal} (the $\sqrt{s}=27$ GeV point is also shown, but the non-monotonicity of the lattice results at $\mu_B\geq 130$ MeV does not allow a determination of $\mu_B$ from it). Lower panel: $R_{12}^Q$ as a function of $\mu_B$.
 The four points correspond to the STAR data for $M_Q/\sigma_Q^2$ at  $\sqrt{s}=27,~39,~62.4,~200$ GeV and centrality 0-10\%, from Ref. \cite{Adamczyk:2014fia}. In both panels, the colored symbols
  correspond to the lattice QCD results in the continuum limit, for the range $(140\leq T_f\leq 150)$ MeV. The arrows show the extracted values for $\mu_B$ at freeze-out.}
\label{fig3}
\end{figure}

\begin{figure}[h]
%\begin{minipage}{0.48\textwidth}
 \scalebox{.7}{
 \includegraphics{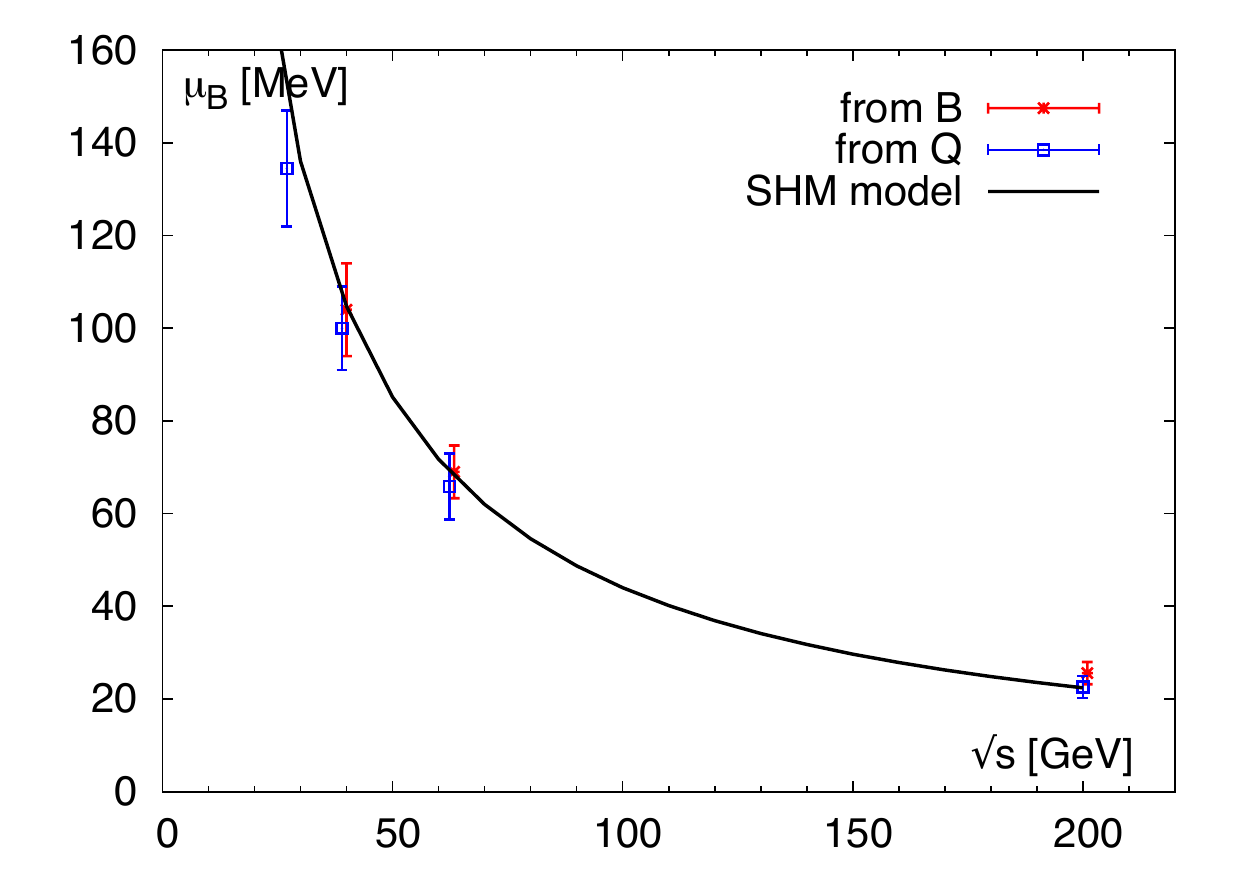}}
% \end{minipage}
%\begin{minipage}{0.48\textwidth}
% \scalebox{.5}{
% \includegraphics{muq}}
% \end{minipage}
\caption{Freeze-out chemical potential $\mu_B$ as a function of the collision energy. The red stars show the $\mu_B$ obtained by fitting $R_{12}^B$, the blue squares have been obtained by fitting $R_{12}^Q$. The black curve comes from the Statistical Hadronization Model analysis of Refs. \cite{Andronic:2005yp,Andronic:2008gu}. \label{fig4}}
\end{figure}

We now proceed to determine the freeze-out chemical potential $\mu_B$, by comparing the lattice results for 
$R_{12}^B$ and $R_{12}^Q$ (as functions of the chemical potential, and in the temperature range $(140\leq T_f\leq 150)$ MeV) to the experimental results for $M_p/\sigma_p^2$ and $M_Q/\sigma_Q^2$ published by the STAR collaboration in Refs. \cite{Adamczyk:2013dal,Adamczyk:2014fia,webpage}. This comparison is shown
in the two panels of Fig. \ref{fig3}: the two quantities allow for an independent determination of $\mu_B$ from electric charge and baryon number: the corresponding values are listed in Table \ref{tab:mu}, and shown in Fig. \ref{fig4}.
Consistency between the two values of baryon-chemical potential is found for all collision energies (the non-monotonicity of the lattice results for $R_{12}^B$ at $\mu_B\geq 130$ MeV does not allow a determination of $\mu_B$ from this observable at $\sqrt{s}=27$ GeV).  Let us now compare the chemical potentials in Table \ref{tab:mu} to those found
earlier in
statistical fits \cite{Cleymans:2005xv,Andronic:2005yp,Andronic:2008gu}. Plotting the
parametrization of Refs.~\cite{Andronic:2005yp,Andronic:2008gu} together with our
values we find a remarkable
agreement (see Fig.~\ref{fig4}). Note that, for the freeze-out temperature,
statistical models typically
yield a somewhat higher value: e.g. 164 MeV in
Refs.~\cite{Adams:2005dq,Andronic:2008gu}.
Towards the lower end in temperature range we find Ref.~\cite{Rafelski:2004dp} with
$T_f=155\pm8$ MeV
with $\mu_B=25\pm 1$ MeV, at $\sqrt{s}=200$~GeV.

 \begin{table}
  \begin{tabular}{|c|c|c|}
 \hline
$\sqrt{s} [GeV]$ & $\mu_B^f$ [MeV] (from $B$)& $\mu_B^f$ [MeV] (from $Q$)\\
\hline
200&25.6$\pm$2.4&22.6$\pm$2.4\\
62.4&69$\pm$5.7&65.9$\pm$7.2\\
39&104$\pm$10&100$\pm$9\\
27& - &134.5$\pm$12.5\\
\hline
 \end{tabular}
 \caption{
\label{tab:mu}
Freeze-out $\mu_B$ vs. $\sqrt{s}$, for the four highest-energy STAR measurements. The $\mu_B$ values and error-bars in this table have been obtained under the assumption that $140$ MeV $\leq T_f\leq150$ MeV. This uncertainty dominates the overall errors. Other (minor) sources of uncertainty are the lattice statistics and the experimental error.}
 %\end{ruledtabular}
 \end{table}

\begin{figure}[h]
\begin{minipage}{0.48\textwidth}
 \scalebox{.7}{
 \includegraphics{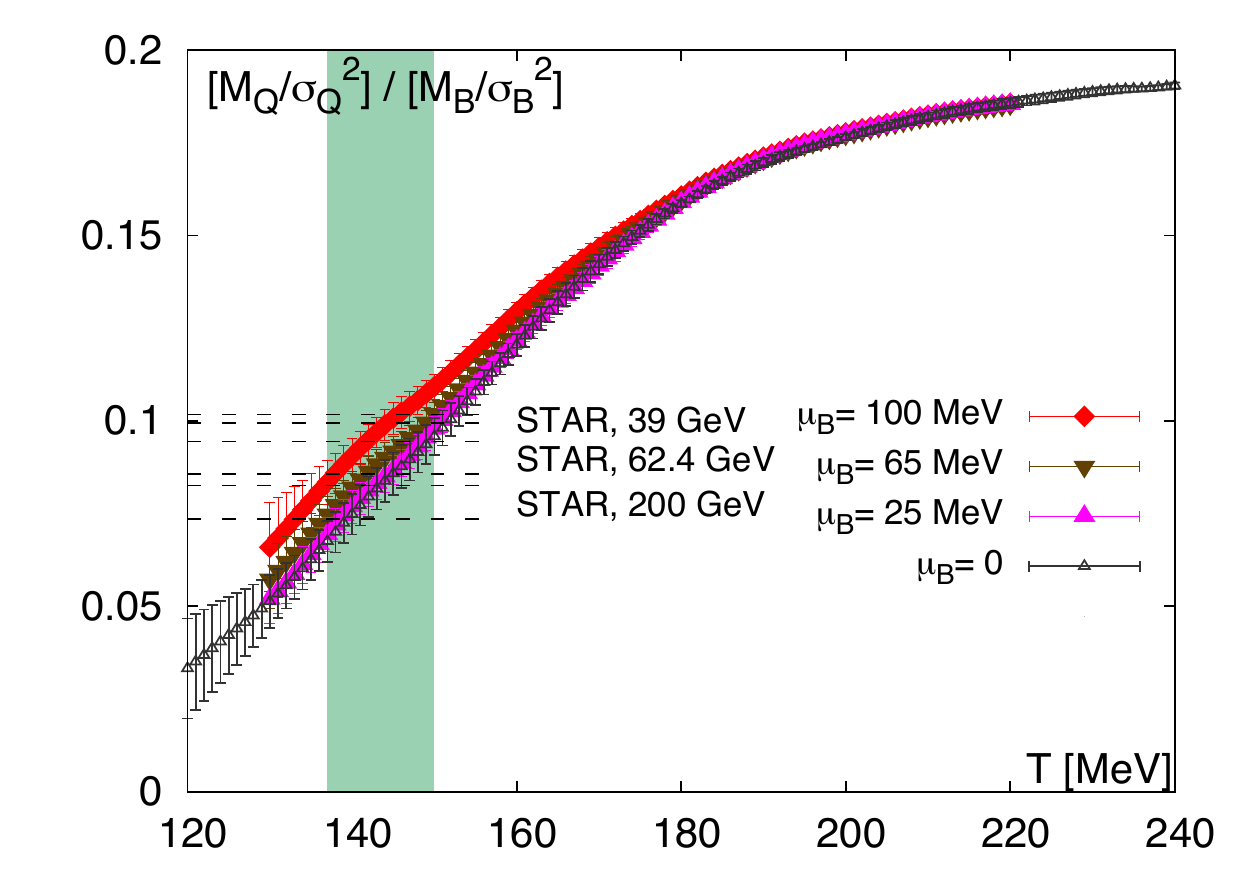}}
 \end{minipage}
\caption{$R_{12}^Q/R_{12}^B$: the colored symbols correspond to continuum-extrapolated lattice QCD simulations at different values of $\mu_B$.
 The dashed lines shows the recent
 experimental measurements by the STAR collaboration \cite{Adamczyk:2013dal,Adamczyk:2014fia} for a 0-10\% centrality and different collision energies. The green-shaded area shows the valid temperature
 range, $T_f=(144\pm6)$ MeV. \label{fig2}}
\end{figure}

The comparison of our lattice results to the latest efficiency-corrected
STAR data hints at a consistency of the freeze-out chemical potential
if we assume an agreement in the temperature. This assumption was well
motivated by the proton and charge skewness data. Let us now take the
assumption further: if the freeze-out can be described by the same
temperature and chemical potentials for charge and protons, then one can
create a combined observable:
$R^Q_{12}/R^B_{12}=[M_Q/\sigma_Q^2]/[M_B/\sigma_B^2]$.  Here, the volume factor
of the charge and baryon (proton) measurements cancel separately.
Should our assumption be correct, this ratio of ratios is the preferable
thermometer: it is far easier to obtain both for lattice and experiment
since it does not involve skewness or kurtosis. We have lattice data available
to $\sim\mu_B^2$ order, which we use when comparing our
results to data. Such a comparison is shown in Fig.~\ref{fig2}.
Contrary to the skewness thermometer, here we see a clear monotonic 
temperature dependence without the hardly controllable lattice errors 
at low temperatures.  This allows for the identification of a narrow
temperature band, instead of an upper limit.

The thermometer in Fig.~\ref{fig2} is, in fact, a consistency criterion:
it agrees to the experimental data at the temperature which needs to
be assumed for the freeze-out if the proton and charge fluctuations
reflect the grand canonical ensemble at a common chemical potential.  For high
enough energies ($\sqrt{s}\ge39$ GeV) this consistency is granted if
freeze-out occurs in the range $T_f=144\pm6$~MeV. Notice that this 
temperature range lies just below the upper limit  that we determined independently
in Fig. \ref{fig1}.

In conclusion, in the present paper we have extracted the freeze-out conditions
(temperature and chemical potential) by comparing our continuum extrapolated
lattice QCD results to the experimental moments of net-charge and net-proton
multiplicity distribution by the STAR collaboration. These new, efficiency
corrected, experimental data point at a lower freeze-out temperature compared
to previous estimates. This is compatible with the expectation that the freeze-out should
occur just below the transition \cite{BraunMunzinger:2003zz}.  The independent
determinations of the freeze-out chemical potentials from electric charge and
baryon number show a remarkable consistency with each other. This comparison is
possible for the first time, and the consistency of the results is of
fundamental importance to validate the hypothesis on which this method is based,
namely that the experimentally created system is close to thermal equilibrium
at the freeze-out and can be described by lattice QCD simulations, at least in
the light quark sector.

\textrm{Acknowledgments:}
C. Ratti acknowledges fruitful discussions with Francesco Becattini, Rene Bellwied and Bill Llope. 
This project was funded by the DFG grant SFB/TR55.
The work of C. Ratti is supported by funds provided by the Italian Ministry of
Education, Universities and Research under the
Firb Research Grant RBFR0814TT.
S. D. Katz is funded by the ERC grant ((FP7/2007-2013)/ERC No 208740)
as well as the "Lend\"ulet" program of the Hungarian Academy of Sciences
((LP2012-44/2012).
The numerical simulations were performed on the QPACE machine, the GPU cluster
at the Wuppertal University and on JUQUEEN (the Blue Gene/Q system of the
Forschungszentrum Juelich).

\bibliography{biblio}

\begin{thebibliography}{117}
\bibitem{webpage}
The efficiency-corrected data for the lowest cumulant ratio ($c_2/c_1$) for net-protons can be found on the public STAR webpage.
\end{thebibliography}
\end{document}